# Magneto-optical methods for magnetoplasmonics in noble metal nanostructures

Alessio Gabbani,[1] Gaia Petrucci,[1] Francesco Pineider[1]*

1 - INSTM and Dipartimento di Chimica e Chimica Industriale, Università di Pisa, Via Giuseppe Moruzzi 13, 56124, Pisa, Italy

*francesco.pineider@unipi.it

**ABSTRACT**

The use of magneto-optical techniques to tune the plasmonic response of nanostructures is a hot topic in active plasmonics, with fascinating implications for several plasmon-based applications and devices. For this emerging field, called magnetoplasmonics, plasmonic nanomaterials with strong optical response to magnetic field are desired, which is generally challenging to achieve with pure noble metals. To overcome this issue, several efforts have been carried out to design and tailor the magneto-optical response of metal nanostructures, mainly by combining plasmonic and magnetic materials in a single nanostructure. In this tutorial we focus our attention on magnetoplasmonic effects in purely plasmonic nanostructures, as they are a valuable model system allowing for an easier rationalization of magnetoplasmonic effects. The most common magneto-optical experimental methods employed to measure these effects are introduced, followed by a review of the major experimental observations that are discussed within the framework of an analytical model developed for the rationalization of magnetoplasmonic effects. Different materials are discussed, from noble metals to novel plasmonic materials, such as heavily doped semiconductors.

## I. INTRODUCTION

The interaction between photons and free electrons in a metal nanostructure can trigger their collective oscillation, which is responsible for the manifestation of fascinating optical properties. This collective motion can either propagate along at the interface between the metal surface and a dielectric, and is then called Surface Plasmon Polariton (SPP), or be confined in a nanostructure, as in the case of Localized Surface Plasmons (LSP). When the energy and momentum of the incident electromagnetic field match the ones of the SPP (or LSP), the system is resonant, resulting in a large extinction of the incoming light together with a strong localization and enhancement of the electromagnetic field in a nano-size region around the metal surface. These two features allows for light manipulation well below the diffraction limit, opening up a wide range of possible applications in the fields of nanophotonics,[1] photocatalysis[2,3] photovoltaics[4] and sensing[5–7] devices.

Among the applications of plasmonics, an important one is represented by surface-enhanced spectroscopies, which exploit the localization and enhancement of the electromagnetic field at the particle surface to increase significantly the spectroscopic signal of a molecule placed closed to the surface. This effect is particularly interesting for surface-enhanced Raman scattering (SERS)[8,9] or surface enhanced luminescence.[10] For instance, Raman spectroscopy is known for giving very low signals, due to the small scattering cross-sections of molecules. Performing Raman spectroscopy on molecules adsorbed on nanostructures can enhance the molecule signal by a factor up to $10^{14}$.[11]

The strong sensitivity of the plasmonic response to the dielectric permittivity of the medium has lead to the development of one the most widespread applications of plasmonics, which is represented by label-free refractometric sensing.[5] This approach allows the detection and quantification of molecular interactions near the surface of the metal by monitoring the shift of the resonance condition induced by local changes of the dielectric permittivity. Within this framework, approaches based on propagating plasmons (SPPs) have been object of intense study, and commercial instruments for biochemical assays and thin film characterization have been developed.[12] The versatility of chemical functionalization makes this approach flexible for several biochemical assays.[13] On the other hand, more recently the advantages of nanoparticle-based plasmonic sensors, such as the higher surface sensitivity and simpler experimental set up were demonstrated.[14,15] Nevertheless, their performances are still lower than the corresponding SPP-based commercial sensors, which opens up a quest for new strategies to enhance the sensitivity of LSP-based refractometric sensor. To this aim, a precise evaluation of the resonance condition is desired, and an interesting perspective is the active modulation of the LSPR response with an external stimulus.[16,17] While the modulation of the spectral position and the magnitude of the LSP resonance (LSPR) is achievable controlling size, shape, dielectric surrounding and plasmonic material of the nanostructure, an active modulation is more challenging and requires more sophisticated strategies. Such an active and reversible control has been achieved through the application of an external stimulus, which can be light itself,[18,19] temperature, an electric field,[20] a chemical stimulus.[21] or a magnetic field.[22]

In this framework, the use of an external magnetic field is of great interest, as it can reversibly modulate the optical properties of the systems at a switching speed that can reach the femtosecond scale. The possibility of exploiting magnetism to modulate the plasmonic response was first demonstrated by Temnov *et al.* by using a hybrid metal-ferromagnet multi-layered structure supporting SPPs.[23] After this seminal work, several combinations of magnetic and plasmonic materials have been exploited to reach an active magneto-plasmonic response in nanostructures.[22,24] Due to their ability manipulate the magnetoplasmonic properties of nanostructures, magneto-optical spectroscopic techniques have triggered significant improvements in the sensitivity of refractometric plasmonic sensing approaches, as reported by Pineider *et al.*[25], Maccaferri *et al.*,[26] and Manera *et al.*.[27]

In addition to using magnetism to modulate the plasmonic response, a second advantage of magnetoplasmonics is the exploitation of plasmon-enhancement to boost the magneto-optical response in hybrid magnetic-plasmonic materials, mediated by magneto-optical coupling between the two counterparts. This is interesting to enhance the performances of optical modulators which are key optical elements in laser and telecommunication devices allowing for non-reciprocal light transmission.[28] Plasmon-enhanced magneto-optics was also demonstrated in hybrid plasmonic-single molecule magnet systems. By putting a molecular layer in direct contact with a plasmonic nanostructure, the magneto-optical response of the molecules can in principle be detected down to the monolayer.[29] This effect can be observed also in hybrid magnetic-plasmonic nanostructures, combining magnetic and the plasmonic materials in a nano-sized object. When combining a magnetic and a plasmonic material, a fair equilibrium must be reached, as the plasmonic material typically displays a sizeable but low magneto-optical response, and need a relatively high magnetic field in order to be recorded (in the order of the Tesla), while on the other hand ferromagnetic metals have considerable optical losses, which result in broad plasmonic resonances.[24]

Several examples of hybrid magnetic-plasmonic materials have been reported, as the magnetoplasmonic nanoantennas made of shape engineered Ni nanodisks prepared by hole-mask colloidal lithography by the groups of A. Dmitriev and P. Vavassori,[30,31] or magnetoplasmonic hybrid materials where the light localization associated with SPRs in gold nanoholes or multilayered Au/Co/Au nanodisks arrays is presented as a way to enhance the TMOKE response through the coupling with an adjacent magneto optical layer.[32] Peculiar spatial arrangements of ferromagnetic (permalloy nanodisks) and plasmonic nanostructures (Au nanorings) were also proposed as innovative approaches to enhance the MO signal through the excitation of dark plasmonic modes,[33] as well as periodic arrangement of Ni or Ni/SiO$_2$/Au nanodisks supporting surface lattice resonance.[34–36]

On the other hand, when dealing with purely plasmonic materials, sharper resonances are generally obtained.[25,37] In this case, the magnetoplasmonic effect is simpler to rationalize, making plasmonic nanostructures suitable model systems to understand the origin of the magnetoplasmonic effects and possibly even foresee the response of more complex structures. Moreover, it was demonstrated that magneto-optical spectroscopies can be employed as a powerful tool to identify the geometry of the plasmonic modes in noble metal nanostructures.[38] These techniques have been also employed to investigate the hybridization between molecular excitonic resonances and the plasmonic resonance of Au@Ag nanorods, which leads to interesting optical and magneto-optical phenomenon in the strong coupling regime.[39,40]

Here we will focus our attention on purely plasmonic nanostructures supporting LSPR (mainly noble metals and heavily doped semiconductors) as model systems to describe the magnetoplasmonic effect. We will first describe the most common magneto-optical effects, detailing their microscopic origin and the main experimental techniques which can be employed for their detection. Magneto-optics will be our technique of choice to experimentally study magnetoplasmonic effects, as it can elucidate the symmetry and the degeneracy of electronic states and plasmonic modes. The advantages of magneto-optical techniques to precisely evaluate the resonance conditions and the effect of the noble metal nanostructure shape on the magneto-optical response will be highlighted within the framework of the theoretical models developed in the literature. Finally, we will discuss the use of heavily doped semiconductors as a promising alternative to noble metals for magnetoplasmonics, and the use of MO techniques to extract free carrier parameters in such plasmonic materials. In this tutorial article, a reader which is familiar with plasmonics concepts can find the basic experimental and theoretical knowledge needed to understand magnetoplasmonic phenomena in nano-sized particles.

## II. FUNDAMENTALS

### II.I POLARIZATION OF LIGHT

The control of light polarization is fundamental in magneto-optical techniques. Therefore, it is first necessary to introduce the concept of polarized light and its interaction with a medium. A plane electromagnetic wave propagating along the z axis can be described according to the following equation:

(1) $E(z,t) = E e^{i(kz-\omega t)}$,

where the electric field of light can be expressed as:

(2) $E = \begin{bmatrix} E_x \\ E_y \end{bmatrix} = \begin{bmatrix} a \\ b e^{i\delta} \end{bmatrix}$.

The above equation describes the electromagnetic wave as a superposition of two waves, one oscillating only along the *x* direction, and the other oscillating only along the *y* direction. In equation 2 *a*, *b* and $\delta$ are real numbers representing respectively the amplitudes of the *x* and *y* components of the electric field and its phase. It can be shown that when $\delta$ is zero, the electromagnetic wave is linearly polarized. In particular, if *b*=0 the polarization plane is along the *x* axis, while for *a*=0 it is along the *y* axis. Other non-null values of *a* ad *b* lead to linearly polarized light having arbitrary orientation of the polarization plane.

On the other hand, when *x* and *y* components of the electric field have a phase delay, i.e. $\delta \neq 0$, the electromagnetic wave is elliptically polarized. The latter means that the electric field traces an elliptic path while it propagates in time or space, as the result of the superposition of the oscillation of the phase-delayed *x* and *y* components. A particular case is represented by circularly polarized light, which occurs when *x* and *y* electric field components have equal amplitude (*a=b*) and a phase delay $\delta$ of $\pm \pi/2$. In the case of $\delta = +\pi/2$ light is defined as right circularly polarized (RCP), while for $\delta = -\pi/2$ it is defined as left circularly polarized (LCP). It follows that CP light can be expressed as a superposition of two phase-delayed linearly polarized waves along the x and y axis respectively, having equal amplitudes. In a similar way it can be shown that linearly polarized light can be expressed as a superposition of two CP waves rotating in opposite directions (LCP and RCP) with no phase delay (Figure 1 A).

To describe the interaction of light with a medium, the complex refractive index should be introduced:

(3) $\underline{n} = n + i\kappa$,

where *n* and $\kappa$ represent the real and imaginary parts of the medium refractive index, which are related to the change in velocity and to the attenuation of the incoming light respectively.[41] When linearly polarized light along the *x* axis interacts with a medium having different refractive index for the two CP components of the incoming light ($n_{RCP} \neq n_{LCP}$), the polarization plane is rotated. If the latter occurs, the phenomenon is called circular birefringence. Moreover, if the medium absorbs differentially one of the two CP components, i.e. the imaginary part of the refractive index is different for RCP and LCP ($\kappa_{RCP} \neq \kappa_{LCP}$), not only the polarization plane is rotated but light becomes elliptically polarized after travelling through the medium. The optical phenomenon of differential absorption of LCP and RCP is called circular dichroism. The resulting elliptically polarized light can be described by defining the two angles $\theta$ and $\psi$ (Figure 1 B): the first represents the *rotation* of the polarization ellipses with respect to the *x* axis, while the second defines the degree of *ellipticity*, i.e. the ratio between the minor and the major axis of the ellipse. The first angle quantifies how much the polarization plane of incoming LP light is rotated with respect to the initial orientation, while the second one defines how much the two CP components of LP light are differentially attenuated. When rotation and ellipticity (or birefringence and dichroism) occur in absence of an applied magnetic field, the medium exhibits *optical activity*, which is the optical manifestation of its chirality.[42] On the other hand, when the magnetic field induces a modification of the optical response of the material, about the effect is defined *magneto-optical activity*, which is the interest of this tutorial article.[43]

Polarization-based techniques are at the basis of the study of chiro-optical and magneto-optical phenomena, and are widely used by chemists, physicists and material scientists. Chemists are more familiar with natural and magnetic circular dichroism, which essentially measure changes in the intensity of RCP or LCP light related to its differential absorption displayed by the sample.[42,44] Such information in magneto-optics is directly related to the electronic structure of the material. On the other hand physicists working in the field of optics are more familiar with ellipsometry spectroscopic techniques,[45] whose magnetic counterpart is represented by Faraday rotation and ellipticity,[43] which give the complete information on the polarization of light after traveling through the sample, i.e. intensity and phase or rotation and ellipticity angles. Despite the fact that the two techniques are often used in different contexts and the physical quantities obtained are expressed in different units (MCD is generally expressed in differential absorption units, while Faraday rotation and ellipticity are expressed in degrees or radians), we want to point out that the physical quantities measured are essentially the same, as we will discuss in the following paragraphs.

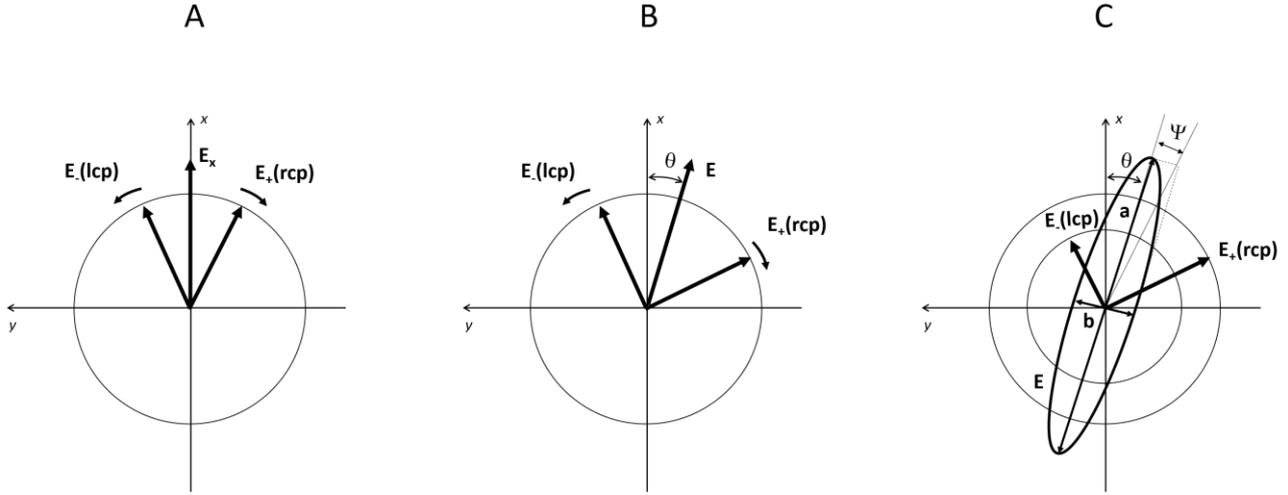

Figure 1: A) Linearly polarized wave described as a sum of two circularly polarized light (CP) with opposite helicity; B) rotation of polarization plane of linearly polarized light: one of the two CP component travels in the medium with a different velocity; C) Rotation and differential absorption of the two CP electric field components (E+(RCP) and E-(LCP)) composing the linearly polarized light, resulting in an elliptically polarized light after the interaction with the medium. a and b are the axis of the ellipses, while the two angles $\vartheta$ and $\varphi$ represent the rotation and ellipticity angles.

**II.II MAGNETO-OPTICAL EFFECTS**

The first experimental observation of magnetic field-induced modification of the optical response of a material dates back to 1845, when Michael Faraday discovered that the polarization plane of light is rotated after the interaction with a magnetized medium. Such magneto-optical phenomenon is called Faraday rotation, or magneto-optical Faraday effect. The angle of rotation is proportional to the applied magnetic field $B$, to the sample thickness ($L$) and to the *Verdet* constant ($V$), which is a material- and frequency-dependent function.

(4) $\theta = LVB$

Faraday rotation is measured in transmission geometry, which in magneto-optics is commonly named as the Faraday configuration, and using linearly polarized light as source, while the applied field is oriented parallel to the light propagation direction (Figure 2 A). Besides Faraday rotation, it is possible to measure the ellipticity angle of the polarization ellipses. From the experimental point of view, the direct measurement of Faraday rotation is sometimes challenging, especially when the matrix or the solvent in which the material of interest is placed constitutes a significant fraction of the total volume of the sample. In this case, the signal of the matrix can mask the signal of the component of interest due to its large volume fraction, according to equation (4), even if it has a relatively low Verdet constant. In this case, it is generally preferred to measure the differential absorption between RCP and LCP light, which is called Magnetic Circular Dichroism (MCD) and is another manifestation of the Faraday effect. Although they are recorded in slightly different experimental conditions, the measured MCD (or ellipticity) and Faraday rotation are the manifestation of the same MO phenomenon, and are connected mathematically through the Kramers Kronig relations,[43] as they are related to the two angles that define the polarization ellipses traced by elliptically polarized light obtained after the propagation of light through the sample. MCD is generally preferred for molecules or nanoparticles dispersed in a transparent matrix, as the Faraday rotation of the latter can be large also in transparent regions of the spectrum. MCD is measured illuminating the sample alternatively with RCP (LCP) light in the presence of a static magnetic field $B$ parallel to the propagation direction (Figure 2 B), and recording the differential absorption $\Delta A$ between the two:[44]

(5) $\Delta A = A_{RCP} - A_{LCP} = \Delta \varepsilon_m CLB$,

where $C$ is the concentration of the molecule or nanoparticle of interest, L is the optical path and $\Delta \varepsilon_m$ is the differential extinction coefficient per unit of magnetic field of the material. The differential absorption measured in MCD can be converted in the ellipticity angle of the polarization ellipse ($\Psi$ in Figure 1 C). Indeed, the tangent of the ellipticity angle is equal to the ratio between the minor and the major axes of the ellipses ($a$ and $b$ in Figure 1 C), which are respectively the difference and the sum of the RCP and LCP electric field component in which light can be decomposed (equation 4).

(6) $\Psi_{rad} \approx \tan\Psi_{rad} = \frac{a}{b} = \frac{E_{LCP} - E_{LCP}}{E_{LCP} + E_{LCP}}$

Substituting the intensity of light to the electric field ($E_{LCP(RCP)} = \sqrt{I_{LCP(RCP)}}$), after some substitutions, equation 6) can be easily rearranged into equation 7):[42]

(7) $\Psi_{rad} \approx tan\Psi_{rad} = \frac{e^{ln10\frac{\Delta A}{4}} - e^{-ln10\frac{\Delta A}{4}}}{e^{ln10\frac{\Delta A}{4}} + e^{-ln10\frac{\Delta A}{4}}} = \tanh\left[ln10\frac{\Delta A}{4}\right] \approx ln10\frac{\Delta A}{4} = 0.57\Delta A$

Exploiting the definition of hyperbolic tangent, and under the assumption of small angles, the well known equation 8) is obtained for the conversion of the MCD signal from $\Delta A$ units into the ellipticity angle in degrees:[42,44]

(8) $\Psi_{deg} = \Psi_{rad} \cdot \frac{360}{2\pi} = \frac{360}{2\pi} \cdot 0.57 \cdot \Delta A = 32.982 \cdot \Delta A$.

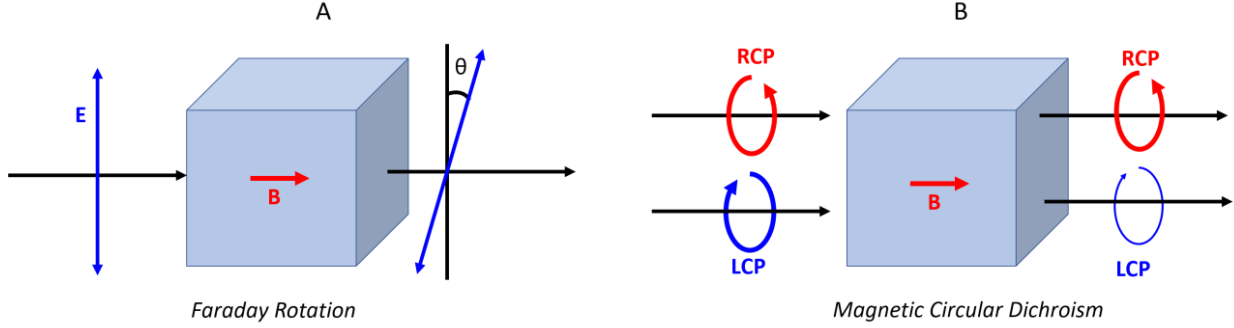

Figure 2: Scheme of Faraday rotation (A) and Magnetic Circular Dichroism (B) effects. In the first case linearly polarized light is used as incoming wave, and the interaction with a magnetized medium causes a rotation of its polarization plane. In the case of MCD, RCP or LCP are alternatively sent to the sample with k-vector parallel to the applied magnetic field, resulting in a differential transmission of the two CP waves.

Within the framework of the macroscopic theory usually employed to describe the magneto-optical phenomena, it is convenient to introduce the notation of dielectric function $\varepsilon = \varepsilon_1 + i\varepsilon_2$, which is often employed instead of $\underline{n}$ (the two quantities can be related through simple equations.[41,46] It is common to use a tensorial form of the dielectric function, where the effect of the magnetic field is generally introduced as an off-diagonal component. For the simplest case of an isotropic medium and working in Faraday configuration with magnetic field oriented along z direction and parallel to the light propagation, the dielectric tensor can be written according to equation 9:

(9) $\begin{pmatrix} \varepsilon & \varepsilon_{xy} & 0 \\ -\varepsilon_{xy} & \varepsilon & 0 \\ 0 & 0 & \varepsilon \end{pmatrix}$,

where each element of the matrix is complex. The off-diagonal component $\varepsilon_{xy}$ is proportional to the magnetization of the sample. Ellipticity ($\Psi$) and Faraday rotation ($\vartheta$) are proportional to the imaginary and to the real part of $\varepsilon_{xy}$, respectively:[47]

(10) $\Psi = -\frac{\omega}{2c} Im\left[\frac{\varepsilon_{xy}}{\sqrt{\varepsilon}}\right]$

(11) $\vartheta = \frac{\omega}{2c} Re\left[\frac{\varepsilon_{xy}}{\sqrt{\varepsilon}}\right]$

An important advantage of MCD spectroscopy over optical spectroscopies is that it provides an in-depth correlation between the MO response and the electronic properties of the investigated materials. In fact, the differential absorption can be correlated to the magnetic field-induced splitting of electronic levels in the material. Thanks to the additional selection rules of the technique (see below), *RCP* (*LCP*) light can excite selectively transitions between energetic levels with the appropriate quantum number $m_j$, going deeper into the electronic structure of the material. Indeed, polarized light of opposite helicity (RCP or LCP) can be rationalized from the point of view of quantum electrodynamics in terms of photons carrying angular momentum of *j*=1 and spin momentum of $m_j$=±1, with the consequence that the transition involving a certain $\Delta m_j$ (+1 or -1) can be excited only by the appropriate helicity (i.e. by a photon with the correct spin momentum).[43]

In the conventional interpretation of MCD spectra, three terms can be defined (Figure 3), named *A*, *B* and *C*, assigned typically to different kinds of transitions.[44] The *A term* is known as the diamagnetic term, as it is temperature-independent. It involves the transition toward excited states whose degeneracy is removed by the applied magnetic field through Zeeman interaction. As sketched in Figure 3 A, the two different excited state levels can be addressed selectively with RCP (LCP) light. As the split energetic states are the empty excited states, the transition probability is equal for both states, leading to a symmetric S-shaped signal centred at the absorption maximum. The field-dependence is linear and the signal is temperature-independent, consistently with the fact that the populations of ground and excited states are not affected by thermal energy. On the other hand, *term C* is typically observed for paramagnetic materials, and is temperature-dependent. In fact, in this case the degeneracy is usually removed from the fundamental state, resulting in two energy levels which can be differently populated depending on the temperature, resulting in different transition probability for the two starting states of the transition (Figure 3 B). For this reason, the spectral fingerprint of *term C* can be a peak or an asymmetric derivative. In this case the field-dependence can reach a saturation, consistently with the paramagnetic nature of the material involved and hysteretic behaviour can be observed in the case of long-range magnetic interaction between paramagnetic centers (i.e. ferromagnetism, ferrimagnetism or anti-ferromagnetism). *B terms* are also typical of diamagnetic materials. They originate from the magnetic field-induced mixing of the zero-field wave-functions (Figure 3 C). Such mixing is typically larger when it occurs between two levels which are close in energy and far from the ground state. In the MCD spectrum, it appears typically as two peaks of opposite sign, centred at the two absorption peaks of the transitions involving the field-mixed excited states. When the energy difference between the levels involved in the mixing is small, it can appear as a "pseudo" *A term*. On the other hand, when the energy distance between the levels involved in the mixing and the ground state is small, it can appear as a couple of "pseudo" *C terms*.

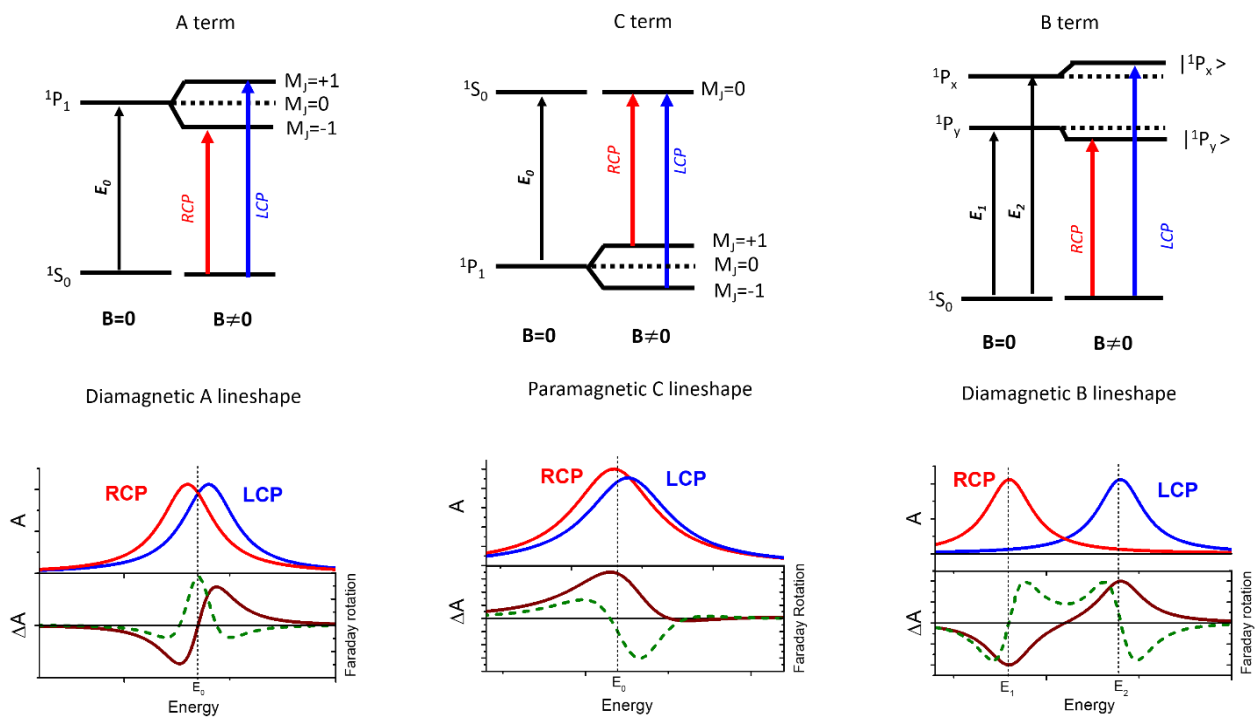

Figure 3: Top panel: scheme of the energy levels involved in MCD transitions (top panel) for the three MCD terms; bottom panel: absorption spectra excited selectively with RCP and LCP light for the transitions shown in the top panel, and resulting MCD signals in differential absorption. The Faraday rotation is represented with green dashed line, bearing the shape of the first derivative of the MCD.

It should be noted that different MCD terms can be present simultaneously in the material, and sometimes variable temperature- or variable magnetic field- measurement are needed to separate the contribution of each term. Moreover, spin-orbit coupling can cause additional splitting of the energy levels in the material, further splitting the electronic levels scheme, increasing the number of MO transitions involved and finally complicating the interpretation of the MCD spectra. The knowledge of the electronic states of the material investigated is predictive about which MCD term is expected. For instance, in the case of molecules or atoms with non-degenerate ground state, no *C term* will be present. Considerations about the symmetry of the molecules or atoms investigated can also be useful:[44] for molecules with a proper or improper rotation axis of order greater than three, excited state degeneracy is possible, resulting in *A terms* for transitions to such degenerate states. On the other hand, B terms are expected for transitions to non-degenerate excited states, even for high symmetry molecules. When the molecules have a rotation axis of two or one, only *B terms* are expected in MCD, as only non-degenerate excited states will be present. Although these rules were traditionally employed for the interpretation of the MCD spectrum of molecules, these simple symmetry considerations can provide useful guidelines also for plasmonic nanostructures,[38] as we will show in the next section.

MO spectroscopy can be a powerful tool to identify different magnetic phases, offering the advantages of combining spectral and magnetic degrees of freedom in the same technique.[48]

Moreover, MO effects can be investigated also in reflection geometry, as discovered by John Kerr in 1877, which first reported the observation of the Magneto-optical Kerr effect, i.e. the rotation of the polarization plane of light after its reflection from a magnetized surface. MOKE spectroscopy is preferable for thick and opaque materials or films, where the amount of transmitted light is not sufficient to measure the Faraday rotation in transmission, such as in the case of ferromagnetic metal films. MOKE spectroscopic technique classification is based on the relative orientation of the applied magnetic field with respect to the plane of incidence of the incoming light (Figure 4): in polar MOKE magnetic field is perpendicular to the sample surface; in longitudinal MOKE the field is applied parallel to the surface; finally in the transversal MOKE, the magnetic field is perpendicular to the plane of incidence. These three methods have the unique advantage of being able to probe different orientations of the magnetization of a magnetic film, which for instance allows extracting the out-of-plane and in-plane magnetization for an ordered ferromagnetic film.[49]

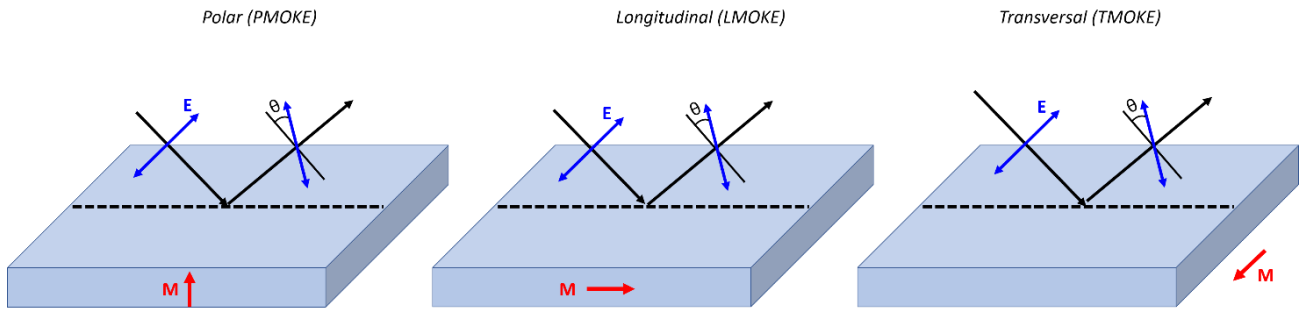

Figure 4: Geometry of the MOKE spectroscopy techniques: in polar MOKE the sample is magnetized perpendicularly to the plane; in longitudinal MOKE in-plane magnetization is investigated, while in transversal MOKE magnetization is applied perpendicularly to the plane of incidence of light.

**II.III EXPERIMENTAL MEASUREMENT OF MAGNETO-OPTICAL EFFECTS**

For the experimental determination of magneto-optical effects, magneto-optical techniques are used. Such techniques require, as all the spectroscopic techniques, a light source, a dispersive element, a light detector and appropriate lenses. In addition, polarization optics and magnetic field source are peculiar elements of magneto-optics, as to observe MO effects an accurate choice of the polarization of light is needed and the sample should be magnetized. Typical experimental set up schemes for the measurement of MCD and Faraday rotation configurations are reported in Figure 5.

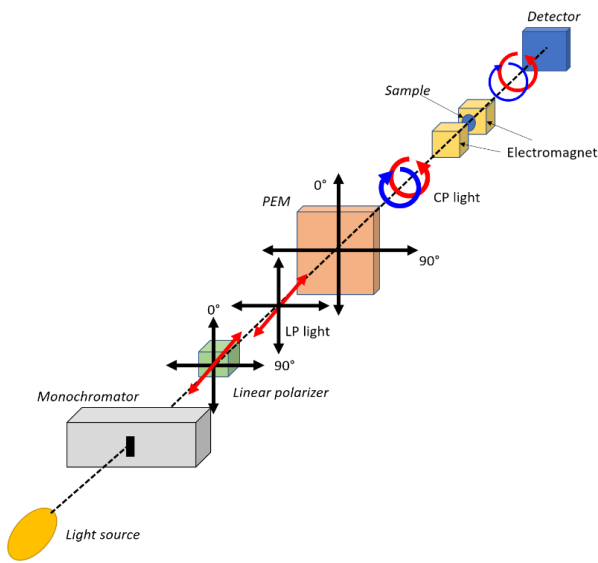

Figure 5: Schetch of a typical Magnetic Circular Dichroism set up.

Since MO effects are typically small (<10$^{-3}$) with respect to optical effects, intense and stable light source are generally preferred, such as high power Xe arc lamps coupled to monochromators, or supercontinuum laser light sources. Polarization optics are used to impart a defined polarization (linear or circular) to light before reaching the sample, or can be used to analyse the polarization state of light after the interaction with the magnetized sample. Often the polarization is modulated between two extreme values (RCP and LCP in MCD spectroscopy) at high frequency in order to improve the sensitivity of the detection strategy, by taking advantage of phase sensitive detection through lock-in amplifiers. Polarization modulation is efficiently obtained with photoelastic modulators (PEMs), which are transparent optical elements able to oscillate at a frequency of the order of tens of kHz along one direction upon an applied voltage due the piezoelectric effect.[50,51] The continuous expansion and compression of the PEM dynamically changes the refractive index of the element along one optical axis, while the orthogonal one is not affected by the oscillation: the material becomes birefringent, i.e. different linear polarization of light travels through the material with different velocity. Thanks to this high frequency modulation, the MO signal can be retrieved and extracted from the total light that reaches the detector by analysing the output signal with a lock-in amplifier. A phase difference is introduced between the two orthogonal components of electric field traveling through the material, called retardation. By properly controlling the retardation of the PEM, one can modulate the polarization. A special case occurs for retardation equal to λ/4: if light is polarized along the bisector of the PEM axes, after travelling through the PEM the output light will oscillate between RCP and LCP light at the PEM frequency (Figure 6). The latter condition is used in the MCD configuration, when ellipticity is measured. On the other hand, when both ellipticity and rotation need to be measured (as in MOKE or Faraday configuration), usually a retardation of 0.383λ is used and an analyser is added before the detector: the signal at the first and second harmonic of the PEM are proportional respectively to ellipticity and rotation.[49,52]

Magnetic field is usually provided by an electromagnet, while for higher field than 1.5 Tesla a superconductive magnet is required using standard optical. Finally, the detection compartment of the set up is equivalent to other spectroscopy techniques: photomultiplier tubes or solid state diode-based detectors are typically used. The detector in MO spectroscopies should have a time-response compatible with PEM modulation frequency.

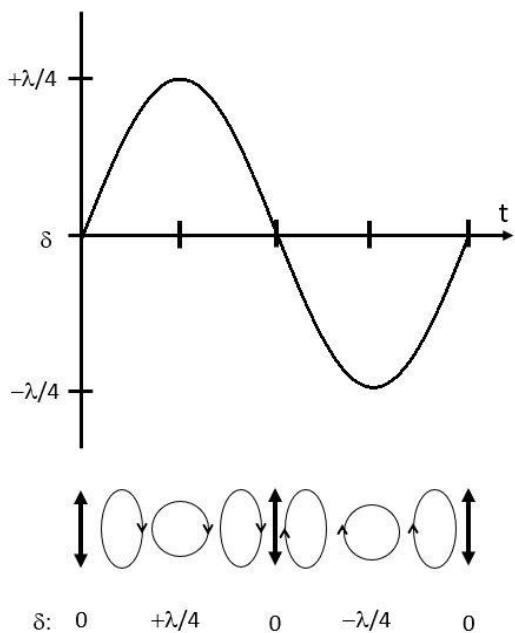

Figure 6: Polarization of light emergent from the PEM over the course of a sin wave modulation. RCP and LCP are produced during a single period.

## III. MAGNETOPLASMONIC EFFECTS IN NON-MAGNETIC PLASMONIC RESONATORS

### III.I ORIGIN OF THE MAGNETIC FIELD INDUCED PLASMON SHIFT

Despite being non-magnetic, noble metals were found to display sizeable magneto-optical properties when they are confined in nanostructures. An early report of the MCD spectrum of 25 nm Au spherical NPs encapsulated in a xerogel or dispersed in water was reported by Zaitoun et al.,[53] who found a temperature independent MCD S-shaped signal, ascribed to an excited state splitting. Since this signal crosses the zero in correspondence to the maximum of the plasmonic resonance in the extinction spectrum (≈ 525 nm), it was ascribed to the LSPR, even if the authors pointed out that the asymmetry of the MCD line-shape suggests further effects in addition to a typical *A term*, which required further studies. Later on, a work by Artemyev et al. reported an MCD investigation on Au, Ag spherical NPs and Au nanorods, describing the effect only phenomenologically.[54] A full rationalization of the experimental MCD response of Au spherical NPs was reported successively by Pineider et al.,[25] on the basis of an analytical model which reproduces the MO response both qualitative and quantitative, applying and extending a theoretical formulation previously reported by Gu and Kornev.[55]

To understand the origin of MO effect in Au NPs, we introduce the Lorentz force acting on each free electron of the metal, according to *Drude* model in the presence of an external magnetic field:

(12) $m\frac{dv}{dt} + \gamma m v = eE + ev \times B$

In the above equation, *m* is the electron mass, $v$ is the electron velocity, *e* the electron charge, *E* represents the electric field of light which is the main force driving the electron oscillation, *B* is the applied magnetic field and $\gamma$ is the damping constant, which slows down the electron oscillation due to electron scattering events. When CP light excites the plasmonic oscillation, one can think of a circular motion of free electrons, which is degenerate for the two helicities in the absence of magnetic field.[56] Nevertheless, when a magnetic field is applied, one should consider the term $ev \times B$ in equation 12, which acts as an additional perturbation to the free electron motion. This additional force experienced by free electrons can be parallel or antiparallel to the primary force due to the electric field of light (*eE* term in equation 12), depending of the direction of $v$ which can be clockwise or anticlockwise depending of the helicity of light (Figure 7 A). The latter results in a different angular frequency of the electron oscillation in the case of RCP (LCP): $\omega_{LCP} \neq \omega_{RCP}$, which is equivalent to say that the degeneracy of circular magnetoplasmonic modes is removed leading to an energy separation between the two. The resonance conditions of a LSPR peak in extinction would thus be equally shifted toward the red or toward the blue by RCP or LCP light. The MCD spectrum resulting from the difference between the two oppositely shifted LSPR peaks have an S-shaped line-shape, similarly to what predicted by MCD theory for a diamagnetic *A term*. This is supported by symmetry considerations, as sphere is highly symmetric, with rotation axis grater than three, and free electron oscillations are degenerate in the incoming electric field *xy* plane.

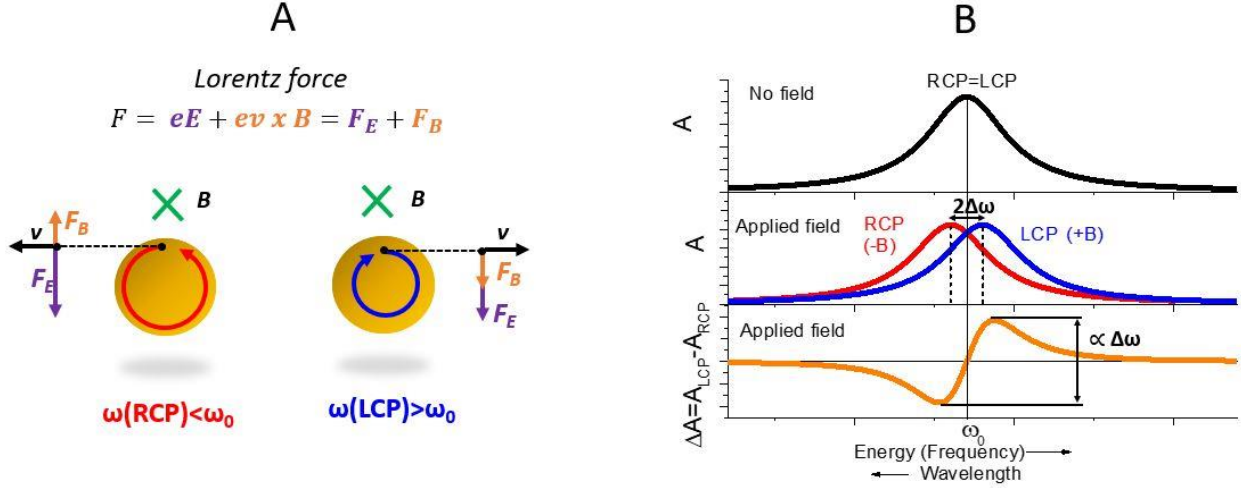

Figure 7: A) Lorentz force acting on each free carrier of the plasmonic resonator in the case of RCP and LCP light in the presence of a static magnetic field. The relative orientation of magnetic and electric component of the force are depicted for RCP and LCP. In this scheme, the electric field of CP light oscillates in the plane of the image, while magnetic field is applied perpendicular to the plane of the image and pointing inside the image plane. B) Extinction spectrum of an ideal plasmonic resonator without the applied magnetic field (black line) centred at the resonance frequency $\omega_0$, extinction spectrum with an applied positive field excited with RCP (red line) or LCP (blue line) light, and resulting differential MCD signal (orange line), crossing the zero at $\omega_0$. The shift of the red and blue peaks with respect to $\omega_0$ is exagerated for an easier understanding of the effect. Given the simmetry of the problem, keeping a fixed polarization (i.e. RCP) but reversing the magnetic field direction (+B or -B) would give the same blue and red curves.

In order to calculate analytically the MO response, it is helpful to recall its relation with the material refractive index. The differential absorption between LCP and RCP light in the presence of an applied magnetic field (MCD) is due to a difference in the imaginary part of the refractive index (*n*) of the material between the two helicities: $MCD \propto Im[n_{RCP} - n_{LCP}]$.[43,44] When dealing with plasmonic nanospheres, instead of using the material refractive index it is more convenient to work with the polarizability of the sphere ($\alpha$), which is a physical quantity often employed in describing the LSPR optical response.[57,58] Within this framework, an expression for the helicity- and magnetic field-dependent polarizability is provided by Gu and Kornev,[55] who applied a perturbation method to solve equation 1, which is reasonable for weak magnetic fields (a few Tesla). For a spherical NP much smaller than the incident wavelength, in the quasi-static dipolar approximation,[58] the field-dependent polarizability can be expressed as follows:

(13) $\alpha_B(\omega) = \frac{\pi D^3}{2} \frac{(\varepsilon(\omega) - \varepsilon_m) + B(f(\omega) - f_l)}{(\varepsilon(\omega) + 2\varepsilon_m) + B(f(\omega) - f_l)}$

In equation 13, *D* is the NP diameter, $\varepsilon(\omega)$ and $\varepsilon_m$ are the dielectric functions of the metal and the medium respectively, while *f(ω)* and *f$_l$* are the coupling functions describing the interaction with the magnetic field for the metal and the solvent respectively. *f(ω)* can be written according to pure *Drude* dielectric function as:

(14) $f(\omega) = \frac{e}{m} \frac{\omega_P^2}{\omega} \frac{(\gamma - i\omega)^2}{(\gamma^2 + \omega^2)^2}$

Due to the spherical symmetry of the NP, changing light helicity (direction of *v* in Figure 7 A) is equivalent to changing the orientation of the applied field *B* from parallel to anti-parallel with respect to light propagation direction, i.e. it has the same impact on the $ev \times B$ term of the Lorentz force. It follows that the MCD response of a metal NP at the magnetic field *B* can be obtained by calculating the polarizability at *+B* and *-B* and the related cross sections according to the following equations:

(15) $\sigma_{\pm B}(\omega) = k\sqrt{\varepsilon_m} imag[\alpha_{\pm B}(\omega)]$

(16) $\Delta\sigma = \sigma_{+B}(\omega) - \sigma_{-B}(\omega)$

For B=0, the standard dipolar polarizability expression is retrieved, with the typical LSPR Frolich condition $(\varepsilon(\omega) + 2\varepsilon_m) = 0$. By writing $\Delta f(\omega) = f(\omega) - f_l = \Delta f_1 + i\Delta f_2$, a generalized Fröhlich condition can be obtained by analytically evaluating the magnetoplasmonic modes $\omega_B$ as a function of the magnetic field intensity and light helicity through a first order series expansion of the dielectric and coupling functions around the LSPR frequency ($\omega_0$).[25] The energy shift can thus be evaluated according to equation 17.

(17) $\Delta\omega = \frac{B\Delta f_1(\omega_0)}{\left|\frac{\delta\varepsilon_1}{\delta\omega}\right|_{\omega_0}}$

Using only the free electron Drude dielectric function (i.e. without including the interband transition contribution) we obtain $eB/2m$, which is equal to half the cyclotron frequency, giving the expression typically associated with the effect of magnetic field on charge carriers:

(18) $\Delta\omega = \frac{eB}{2m} = \omega_C/2.$

By using the full experimental dielectric function for Au (from Johnson and Christy[59]) with correction for the NP size[57] and applying equations 13-16, an excellent agreement with the experiment was obtained for Au NPs (Figure 8).[25] It should be noted that in Au NPs the presence of interband transitions close to the LSPR resonance (521 nm) causes an asymmetry in the MCD signal, with the positive lobe having lower intensity, as it is closer to such sources of optical losses. To reproduce this asymmetry the use of the full dielectric function in the calculation is necessary. Such asymmetry is significantly reduced shifting the LSPR far from the interband transitions of Au, which was achieved by Gabbani et al. by coating the Au core with a dielectric shell.[60] Such dielectric coating with a non-magnetic ultrathin $FeO_x$ layer was also found to enhance the peak-to-peak MCD signal by a factor of 18% and 50 % respectively. Moreover, the energy shift obtained through the fitting of the MCD spectra was in excellent agreement with the analytical expression of $\Delta\omega$ as a function of the resonance energy (calculated through equation (17)), and it was found to significantly increase as a function of the dielectric coating. These findings provide an interesting approach to control and enhance the magnetic modulation of LSPR without adding lossy magnetic materials.

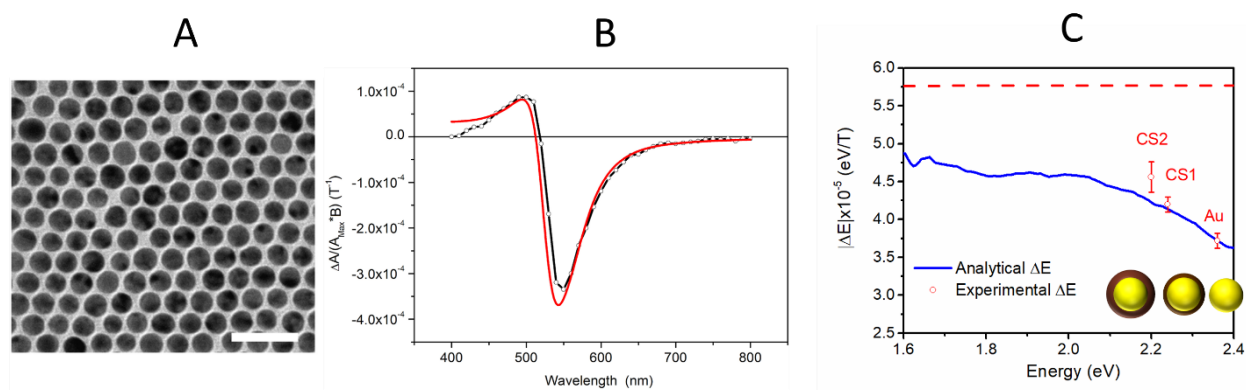

Figure 8: (A) TEM image of 13 nm Au NPs; (B) Experimental (black line) and calculated (red line) MCD spectrum for the Au NPs in (A). Calculated MCD is normalized for the extinction maximum and for the applied magnetic field. Similarly the differential cross section for 1 NP is normalized for the NP extinction coefficient; adapted with permission from reference 25; (C) xperimental and analytically calculated energy shift for Au nanospheres, and core@shell Au nanospheres with non-magnetic FeOx shell of 0.5 (CS1) and 1 nm (CS2). Adapted with permission from reference 60.

A similar derivative-like shape in the ellipticity spectrum was reported by Sepulveda et al. on Au nanodisks embedded in a dielectric medium using polar MOKE spectroscopy (Figure 9 A).[37] The MO response was explained in terms of a Lorentz force which induces a tilt in the oscillating dipole by an angle $\vartheta$ with respect to the polarization plane of incident light (Figure 9 C). Similar results were obtained by our group on Au nanodisks of similar size and aspect ratio prepared by colloidal lithography and analysed through MCD at normal incidence with the substrate (Figure 9 E).[29] Indeed, similarly to the case of the sphere, the symmetry required to obtain magnetic-field induced splitting of the circular magnetoplasmonic modes is maintained in the xy plane, as the disks are oriented perpendicularly to the light propagation direction. With respect to the case of the sphere, increasing the aspect ratio of the disks, the LSPR position is shifted far from the interband optical transitions of Au, leading to an almost symmetric lineshape due to the lower overlap between LSPR and interband transitions. As reported in Figure 9 A, the asymmetry of the ellipticity (MCD) S-shaped signal decreases by shifting the extinction peaks toward the red, and the intensity of the MO signal can be enhanced by increasing the aspect ratio from 2.2 to 6. The magnitude of the magnetic modulation can be extracted by the simultaneous fitting of the extinction and MCD spectra (Figure 9 D-E). To this purpose, first the extinction is fitted with a typical Lorentz function for the LSPR and one or more gaussian functions for the interband components. Then, the MCD is fitted with the difference between two Lorentz functions having the Area and the width fixed at the ones used to fit the LSPR extinction component, but oppositely shifted in position by a quantity $\pm\Delta\omega$ with respect to the extinction maximum $\omega_0$, exploiting an approach previously reported to fit the MCD of quantum dots and dilute magnetic quantum dots.[61,62] The value of $\Delta\omega$ retrieved from the fitting is the magnetic field-induced frequency shift, which depends linearly on the applied magnetic field for non-magnetic plasmonic nanostructures. An excellent agreement was obtained with the experimental spectrum for Au nanodisks, retrieving an energy shift of 0.022 meV/Tesla (Figure 9 D-E).

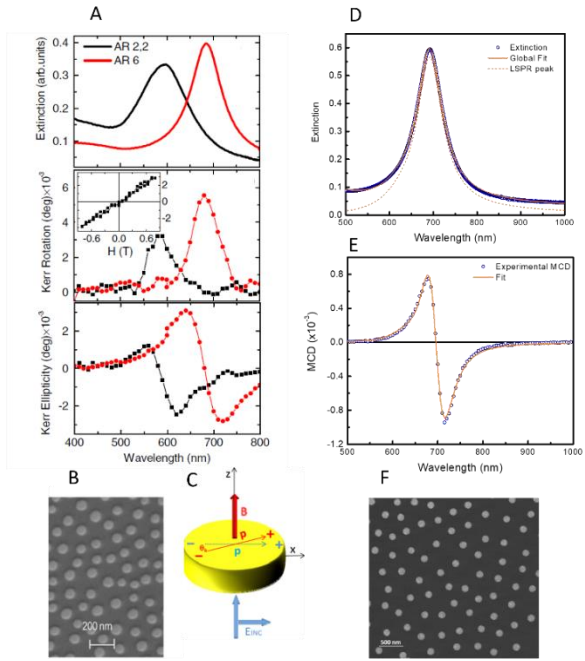

Figure 9: A) From the top to the bottom: extinction, Kerr rotation and Kerr ellipticity at 0.8 Tesla for two nanodisks with aspect ratio of 2.2 (black) and 6 (red); B) SEM image of Au nanodisks analyzed in A; C) Schematic of the MO effect induced by the Lorentz force in a metal nanoparticle; D) extinction and E) MCD spectra at 5 Tesla of Au nanodisks, and fitting of the two spectra; F) SEM images of the nanodisks analyzed in D-E. Adapted from references 29,37

In a recent work, Pedersen applied a semianalytical perturbative approach to calculate the energy shift for Drude metal nanostructures of different shapes.[63] A smaller energy shift was obtained for nanodisks with respect to nanospheres and nanoellipsoids, and was rationalized in terms of local surface electric field components. Since the magnetic field effect induced a modification of the electron motion which is always directed along the xy plane (for k-vector along z), local field components along z will not be affected by the magnetic field. It follows that nanoparticles with the local field completely lying in the xy plane will display a larger magnetoplasmonic effect. While the latter occurs for spheres, for nanodisks a significant component along z is present near the corners, thus decreasing the magnetic field induced energy shift. These results pointed out that the sphere is the optimal shape for magnetoplasmonics.

In order to increase the magnitude of the MCD signal, plasmonic resonators with sharp LSPR peaks are preferred. In fact, as shown in Figure 9 A, since the magnetic modulation acts shifting the resonance toward lower (higher) energy in the case of LCP (RCP), sharper peaks results in larger MO signals, increasing the slope of the signal in correspondence to the plasmonic resonance. A confirmation of the latter is represented by the comparison of the MCD response of Au and Ag NPs (analytical calculations are displayed in Figure 10 A-B), in which the MCD signal is found to be considerably larger for Ag, even though the magnetic field-induced shift $\Delta\omega$ is comparable, as they have similar cyclotron frequency.

Nevertheless, Ag displays a lower LSPR bandwidth due to the lower overlapping of LSPR and interband transitions, which are located at higher energy for silver. This leads to a sharper extinction peak which also results in a larger MCD signal. The influence of the LSPR line-width on the MCD signal at a fixed $\Delta\omega$ is depicted in Figure 10 C-D. Experimental observation of this were reported in the work of Artemyev et al.,[54] and more recently in the work by Yao and Shiratzu,[64] even if the effect was not rationalized quantitatively. A confirmation of the importance of inter-band transition in determining the asymmetry of the MCD signal was also provided by Kovalenko et al. through a comparison between Au and Ag NPs.[65] The authors also reported that inter-particle coupling in 2D networks of Au NPs can significantly alter the MCD line shape, red shifting the LSPR and reducing the asymmetry of the signal.

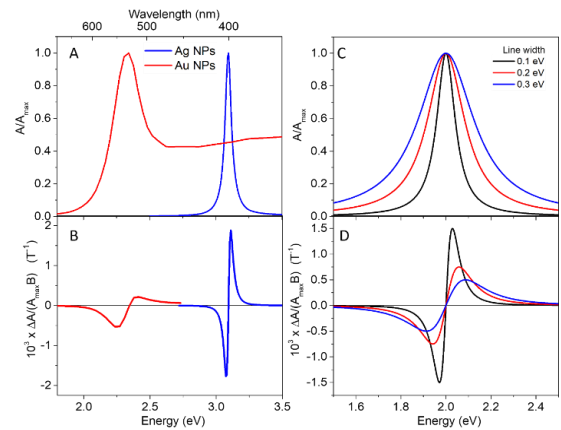

Figure 10: A-B) Analitical calculation of the normalized extinction cross section (A) and the normalized MCD spectrum (B) of Au (red) and Ag (blue) NPs with a diameter of 10 nm. Equations (13)-(16) are used for the calculations, employing the experimental bulk dielectric function of Au and Ag from Johnson and Christy.[59] Both the calculated extinction and MCD are normalized for the extinction cross section maximum. C-D) Extinction and MCD response of Lorentzian resonators with different line width, but same area and position of the resonance peak and same magnetic field-induced shift $\Delta\omega$ are depicted to show the effect of LSPR line width on the MCD response. Sharper resonances gives larger MCD peak-to-peak signals.

### III.II EFFECT OF SHAPE ANISOTROPY

The model described in the previous paragraph is not applicable when the rotational symmetry is broken in the xy plane. This is for instance the case of Au nanorods, which were reported to have a significantly distinct MCD spectrum with respect to Au spheres due to the lower symmetry. After the first observation of the MCD response in plasmonic nanorods,[54] the successive work by Melnikau et al.[40] elucidated the relation between the geometry of the nanostructure and the MO response in chemically prepared nanorods by using qualitatively the formalism based on the MCD terms, while the full quantitative rationalization was developed by Han et al..[66] Transversal and longitudinal LSPR were found to exhibit a significantly different MCD line shape: in the case of transversal LSPR, an S-shaped signal similar to the Au sphere case is displayed, while for the longitudinal resonance, a negative MCD peak is observed. The different response of the two modes were rationalized within the framework of *A* and *B* terms, as no ground state degeneracy is expected in this system. Due to the lower symmetry than the sphere, two LSPR resonances can be excited, one along the transversal axis (TSPR) and one along the longitudinal axis (LSPR) of the rod (Figure 11 A-B). The two modes can be represented as two transitions toward two different excited states. Depending on the symmetry of the modes, we can expect *A* or *B* terms in MCD. Since the longitudinal mode has a 2-fold rotational axis, it is non-degenerate. On the other hand, the transverse mode has a high symmetry as it has a $C_\infty$ axis represented by rotation around the longitudinal axis of the rods (Figure 11 A), and is thus degenerate. According to the latter consideration, an S-shaped signal consistent with an *A term* is observed when the 2-fold degenerate TSPR mode is excited. In addition, according to what predicted by MCD theory, a *B term* can arise from the magnetic field-induced mixing of two excited states, when they are sufficiently close in energy, as in the case of TSPR and LSPR. This is manifested with two peaks of opposite sign, one close to the TSP wavelength and the other at the LSP resonance. The higher is the aspect ratio, i.e. the higher is the separation between the two modes, the lower is the intensity of the B term, as shown in Figure 11 H.

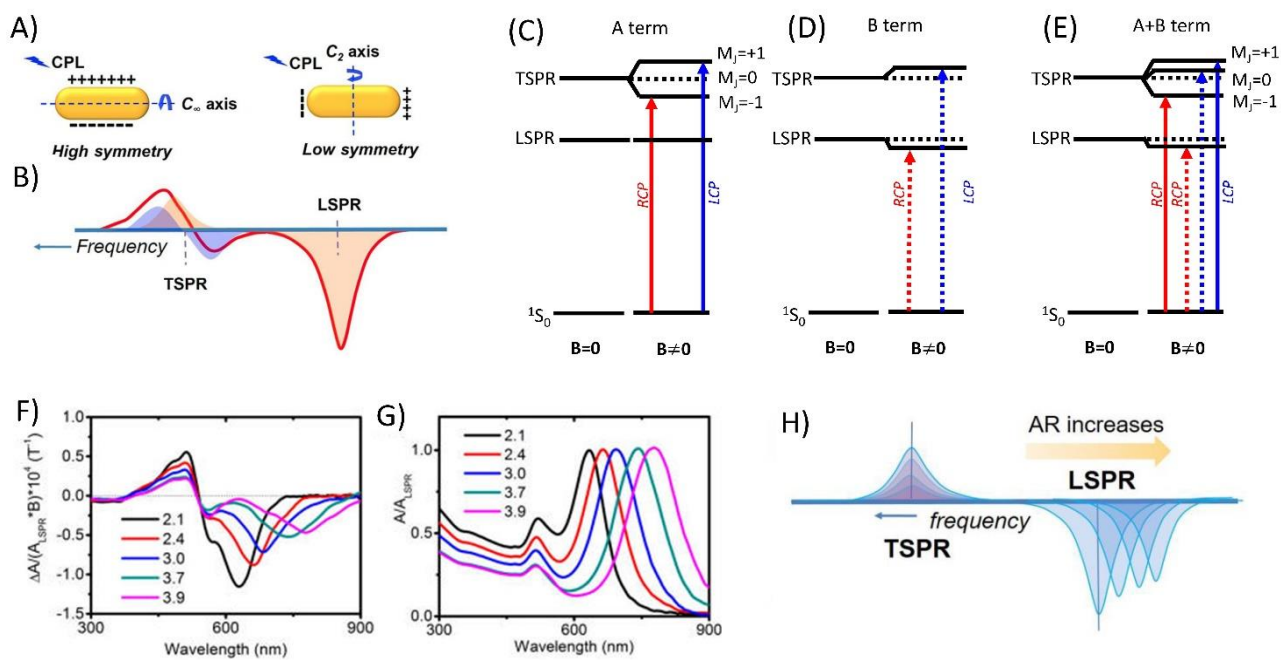

Figure 11: A) Symmetry of transversal and longitudinal LSPR modes in nanorods, and sketch of the MCD signals of the two modes (B). The high energy derivative-like signal (blue-shaded) correspond to a typical MCD A term (energy levels involved are displayed in C), while the other two peaks with opposite sign correspond to a typical B term (energy levels are represented in D). The sum of the two signals gives the overal MCD spectrum (energy levels depicted in E). F-G): sketch of extinction and MCD spectra of Au nanorods of different aspect ratio (from 2.1 to 3.9). H) sketch of the effect of aspect ratio on the MCD of Au nanorods due to the B term. Adapted with permission from ref. 66.

A similar spectral line shape of MCD was found also in randomly oriented Ag nanodisks, prepared by chemical etching by Kovalenko *et al.*.[65] In contrast with the work on Au nanodisks assembled on glass discussed previously, where the electric field of light can only excite the in-plane LSPR mode, in randomly oriented nanodisks both the transversal and the longitudinal LSPR modes are excited, giving a combination of A and B terms, similarly to the case of Au nanorods. Interestingly, in this work the authors gave an interpretation of the MCD on the basis of k-dependent effects related to LSPR intraband transitions within the framework of the band structure of Ag.

Besides nanorods, the extension of this complete model to plasmonic nanostructures of more complex shapes has not been reported yet, even though a rationalization of the response has been provided in a few cases. Work by Yao and Shiratzu reported the MCD response of silver nanoprisms[67] and nanocubes[68] of variable size. In such nanostructures the symmetry is lower than the sphere, and different modes can be excited. For instance, in Ag nanoprisms three LSPR modes are presents, i.e. the dipolar, the in-plane and out-of-plane quadrupolar resonance. These different modes in MCD give an A term, which corresponds to the in-plane dipolar LSPR (with 3-fold rotation symmetry), and a B term corresponding to in-plane or out-of-plane quadrupolar LSPR (without rotation simmetry). The A term is usually the larger contribution, but the B term appears with a typical sharp feature at higher energy. A clear indication of the absence of a pure A term is also suggested by the fact that the signal does not cross zero in correspondence to the extinction maximum. In Ag nanocubes a derivative-like shape centred close to the extinction maximum is detected, similarly to the spherical NP case, while sharp positive peaks are found in the spectrum in correspondence to the two low symmetry modes typically appearing for cubes at higher energy. Nevertheless, a full rationalization of the MCD spectrum of nanocubes and nanoprisms still requires further investigations.

**III.III HEAVILY DOPED SEMICONDUCTORS**

In the last decade, the discovery of LSPR in heavily doped semiconductors has extended the field of plasmonic materials beyond traditional noble metals. In semiconductors, electron doping can be achieved through several strategies, such as the insertion of aliovalent atoms in semiconductor crystals (Sn in $In_2O_3$ or In in CdO), creating vacancies (such as in $Cu_{2-x}S$) or can be induced through post-synthetic photo-treatment.[69,70] By controlling the doping process, the carrier density can be tuned over a wide spectral range, spanning from the NIR toward the far infrared, providing an additional degree of tunability with respect to conventional noble metals, where only shape and size can be tuned to control the spectral position of LSPR. Among heavily doped semiconductors, transparent conductive oxides are particularly interesting as they display relatively low carrier densities and are thus transparent to visible light, having a wide spectral separation between LSPR and interband transitions. Due to the latter, these NPs exhibit sharp LSPR peaks in the near infrared region of the spectrum. However, there is more in these new plasmonic materials that makes them attractive for the field of magnetoplasmonics. Indeed, they can have a significantly lower electron effective mass with respect to noble metals, which is an interesting feature to increase the separation between circular magnetoplasmonic modes, i.e. the magnetic modulation of LSPR which originates the MCD signal. In fact, according to equation 18, for pure Drude-like spherical resonators the magnetic field induced energy shift is inversely proportional to the free carrier effective mass,[25] predicting a significantly higher magnetic modulation in the case for instance of indium tin oxide (ITO) NPs, in which the electron effective mass is typically $0.3$-$0.4 m_e$, where $m_e$ ($9.11 \times 10^{-31}$ Kg) is the mass of free electrons. A pioneering work by the Gamelin group reported the MCD spectra of three classes of heavily doped semiconductors in the near infrared: ZnO, ITO and $Cu_{2-x}Se$ spherical NPs, which were compared to Ag and Au nanostructures (Figure 12).[71] In particular, MCD response of ITO and ZnO NPs were found to be almost one order of magnitude greater than Ag and Au NPs. This was ascribed to a lower value of the electron effective mass which was found to be $0.32 m_e$. Interestingly, since in $Cu_{2-x}Se$ NPs the doping is p-type, and the magnetic field induced energy shift is proportional to the charge of the carrier (equation 18), the MCD signal is inverted in sign for this semiconductor NPs. Indeed, MCD was revealed to be a promising tool to extract the electron effective mass of these class of plasmonic semiconductor NPs, as well as the sign of the charge carriers. A further advantage of transparent conductive oxide NPs is represented by the possibility to co-dope them with paramagnetic ions, opening interesting perspectives toward the preparation of hybrid materials where a first dopant (i.e. Sn in ITO) provides free electrons, while a second one introduces magnetic properties (i.e. *3d* magnetic ions such as iron). To this aim, a careful evaluation of the magnetic doping level is required as impurity ions generally increase electron scattering and creates traps for free electrons decreasing the carrier density, which makes the preparation of magnetoplasmonic hybrid heavily-doped semiconductors NPs rather challenging. Preliminary attempts to couple plasmonic properties and magnetism in these NPs were successfully achieved by Nag's group by co-doping ITO NPs with Fe- and Mn- ions.[72,73] As proven by EPR spectroscopy, free electrons provided by Sn(IV) dopants were found to mediate the interaction between paramagnetic centres (Fe or Mn co-dopants) leading to ferromagnetism, which is maintained at room temperature for the case of iron doping. Such exchange interaction can be particularly interesting for spintronics and magneto-optics, as potentially electron transport in ITO can be controlled by magnetism and the magnetoplasmonic response can be enhanced by coupling free electrons with the spins of the paramagnetic co-dopant. However, to date no direct influence of the paramagnetic dopant in the magnetic modulation of LSPR has been reported.

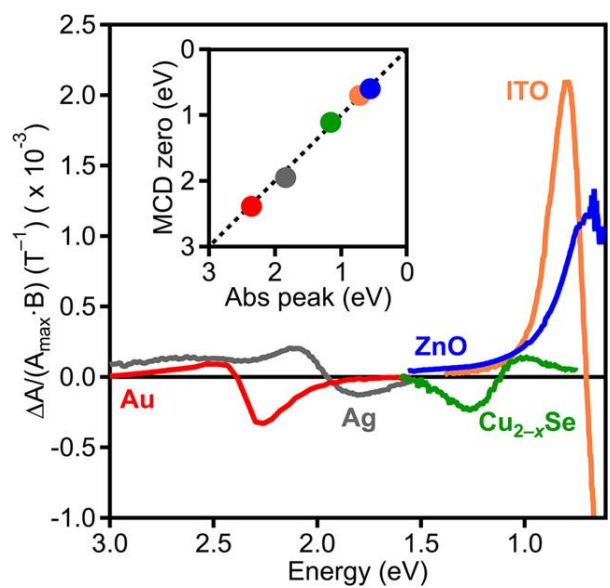

Figure 12: Magnetic Circular Dichroism of three semiconductor NPs compared to noble metals nanostructures. Adapted with permission from reference 71.

Another interesting application of doped semiconductors could be represented by spintronics, especially for magnetic dilute semiconductors. To this aim spin polarization of charge carrier is desired. Within this framework a possible role of magnetoplasmonic modes supported by semiconductors in the spin polarization of charge carrier was reported by a recent study by Radovanovic et al..[75] The authors reported the emergence of a negative peak in the MCD signal in the exciton region (at the semiconductor band gap energy) for ITO NPs which was ascribed to a non-resonant plasmon-exciton coupling. The latter was supported by the fact that in pure $In_2O_3$ NPs, in the absence of Sn- dopant, no MCD signal was observed in the band gap region. A further confirmation that the effect is related to the plasmon excitation was found in the temperature independent nature of the signal, together with its linear field-dependence. The application of a magnetic field removed the degeneracy between circular magnetoplasmonic modes that are excited by RCP (LCP) light. The magnetoplasmonic modes are supposed to couple to the exciton transferring the angular momentum to the conduction band states. Moreover, the blue shift of LSPR with dopant concentration is followed by a blue shift of the MCD peak at the band gap. Splitting of the band states and selective carrier polarization was manipulated further with spin–orbit coupling introducing paramagnetic $Mo^{5+}$ dopant cations instead of Sn. In a successive work, the same authors reported that the band splitting has two contributions: a temperature-independent one arising from the excitation of the circular magnetoplasmonics modes, and a temperature-dependent one, due to localized electron spins associated with defects trap states.[76] The ratio between the two mechanisms can be controlled by tuning the aliovalent doping with Sn or by controlling the formation of oxygen vacancies in $In_2O_3$. The full rationalization of mechanism of this coupling is still unclear and needs further investigations. However, further studies on the correlation between magnetoplasmonic modes and spin polarization of charge carriers can potentially trigger new opportunities both in the field of magnetoplasmonics and spintronics, with potential implications for future applications in information technology.[77]

**IV. CONCLUSIONS AND FUTURE PERSPECTIVES**

In this tutorial article, the main observations concerning magnetoplasmonic effects in purely plasmonic materials were reported. Despite the fact that purely plasmonic materials have lower magneto-optical signal with respect to ferromagnetic and hybrid magnetic-plasmonic systems, their MO response can be easily measured by taking advantage of polarization modulation and phase-sensitive techniques. Moreover, the rationalization of their MO response can be more easily achieved by using an analytical approach based on field-dependent quasi-static polarizability and the excitation of circular magnetoplasmonic modes, which allows precisely extracting the magnitude of the magnetic modulation of LSPR. The MO response can also provide insight into the symmetry of the plasmonic modes, and it is strongly dependent on the nanostructure orientation with respect to the polarization plane of incoming light for anisotropic nanoparticles. The established understanding and rationalization of these peculiar features of the MO response of plasmonic nanostructures represents the basis for the modelling of hybrid nanostructures where the plasmonic material is coupled with molecular resonances or magneto-optically active materials, which confer more advanced functionalities that can be exploited for active magnetoplasmonic devices, such as refractometric sensors with superior performance with respect to the purely plasmonic ones.

The recent discovery of magnetoplasmonic effects in heavily doped semiconductors has opened up the field to new materials, where this effect is considerably increased with respect to noble metals by at least one order of magnitude. This makes the MO response of plasmonic nanostructures not far from being competitive with ferromagnetic and hybrid metal-ferromagnetic nanostructures, with the additional advantage of featuring sharp LSPR. The most promising strategies to boost the magnetoplasmonics response in such nanostructures involve: (i) an accurate dopant choice, (ii) the minimization of electron scattering mechanisms, (iii) the choice of materials with low carrier effective mass, and (iv) the choice of appropriate magnetic co-dopant which couples to free electrons. Despite heavily doped semiconductors representing a promising material for magnetoplasmonics, their application in this field was not intensively explored so far, and by exploiting the recent advancements in colloidal synthesis and dopant engineering, further innovations in this field are potentially feasible employing this novel class of plasmonic materials.

**ACKNOWLEDGMENTS**

Authors acknowledge the financial support of H2020-FETOPEN-2016-2017 Grant No. 737709 FEMTOTERABYTE (EC), PRIN2017 Grant No. 2017CR5WCH Q-ChiSS (Italian MIUR) and of PRA_2017_25 (Università di Pisa).

**DATA AVAILABILITY**

The data that support the findings of this study are available within the article.